\documentclass[fleqn,usenatbib]{mnras}

\usepackage{newtxtext,newtxmath}
\usepackage{makecell}
\usepackage{nicefrac}
\usepackage{threeparttable}

\usepackage[T1]{fontenc}

\DeclareRobustCommand{\VAN}[3]{#2}
\let\VANthebibliography\thebibliography
\def\thebibliography{\DeclareRobustCommand{\VAN}[3]{##3}\VANthebibliography}

\usepackage{graphicx}	
\usepackage{amsmath}	
\usepackage{caption}
\usepackage{subcaption}

\newcommand{\src}{XTE~J1701$-$462}

\title[mHz QPOs in \src \;]{Discovery of millihertz Quasi-Periodic Oscillations in the Low Mass X-Ray Binary XTE~J1701$-$462 from a Search of the \textsl{RXTE} Legacy data set}

\author[K. Tse et al.]{
Kaho Tse,$^{1-3}$\thanks{E-mail: kaho.tse@monash.edu}
Duncan K. Galloway$^{1-4}$
and Alexander Heger$^{1-3,5}$,
\\
$^1$School of Physics and Astronomy, Monash University, Victoria 3800, Australia\\
$^2$OzGrav-Monash, School of Physics and Astronomy, Monash University, VIC 3800, Australia\\
$^3$Joint Institute for Nuclear Astrophysics - Center for the Evolution of the Elements (JINA-CEE), Monash University, Vic 3800, Australia\\
$^4$Institute for Globally Distributed Open Research and Education (IGDORE)\\
$^5$ARC Center of Excellence for Astrophysics in Three Dimensions (ASTRO-3D), Australia\\
}

\date{Accepted XXX. Received YYY; in original form ZZZ}

\pubyear{2023}

\begin{document}
\label{firstpage}
\pagerange{\pageref{firstpage}--\pageref{lastpage}}
\maketitle

\begin{abstract}
We report the detection of millihertz quasi-periodic oscillations ($\mathrm{mHz}$ QPOs) from the low-mass X-ray binary \src.  The discovery came from a search of the legacy data set of the {\it Rossi X-ray Timing Explorer}, in order to detect the periodic signals in all observations of sources exhibiting thermonuclear bursts.  We found that $47$ out of $860$ observations of \src \; covering the 2006--7 outburst exhibits signals with a significance above the detection threshold, which was determined separately for each observation via a Monte Carlo approach.  We chose the four strongest candidates, each with maximum power exceeding $4\sigma$ of the simulated wavelet noise power distribution, to demonstrate the properties of the QPOs.  The frequencies of the signals in the four observations are $\sim 3.5\;\text{to}\;5.6\; \mathrm{mHz}$, and the fractional R.M.S.\ amplitudes vary between $0.74 \pm 0.05\,\%$ and $3.54 \pm 0.04\,\%$.  Although previously reported signals in other sources typically disappear immediately before a burst, we do not observe this behaviour in \src.  Instead, we found that the QPOs and bursts occurred in separate accretion regimes.  When the persistent luminosity dropped near the end of the outburst, the source showed bursts and no QPOs were detected, which is the behaviour predicted by theory for the transition from stable to unstable burning.  On the basis of this new detection, we reassess the cases for identifying these $\mathrm{mHz}$ QPOs in this and other sources as arising from marginally stable burning.


\end{abstract}




\section{Introduction}
The discovery of Type I X-ray bursts from low mass X-ray binaries (LMXBs) in the 1970s \citep{Grindlay1976} has opened up a new window to study nuclear physics in extreme environments.  These flashes are caused by explosive nuclear burning on the surface of a neutron star in a binary star system.  Typically, a mixture of hydrogen and helium is accreted from a donor star onto the neutron star, and a burst can be triggered once the accumulated layer gets hot and dense enough \citep{lamblamb}.  Depending on the hydrogen mass fraction of the accumulated layer, Type I X-ray bursts usually last for tens to a hundred seconds.  For a burst with more hydrogen present, the burst tend to last longer as the burning timescale is longer for H burning \citep{bildsten2000,Woosley2004}.

X-ray bursts have drawn a lot of attention from observers for their utility as a laboratory to test the theory of unstable nuclear burning.  Most X-ray telescopes have observed these X-ray flashes.  The Proportional Counter Array (PCA \citealt{pca}) instrument onboard the \textsl{Rossi X-ray Timing Explorer (RXTE)} satellite has captured more than 1,000 Type I bursts in 10 years \citep{galloway2008,minbarpaper}.  Combining those observed by \textsl{BeppoSAX} and \textsl{INTEGRAL}, 7,083 bursts from 85 bursting sources detected from about 1996 to 2012 have been assembled and studied in detail by \cite{minbarpaper}.

As the number of observations continues to grow, more new features from Type I X-ray bursts have been discovered.  For example, \cite{firstqpo} found quasi-periodic oscillations at milliHertz frequencies in sources 4U~1636$-$536, 4U~1608$-$52 as well as Aql~X-1.  The oscillations drift between 7--9 $\mathrm{mHz}$ with R.M.S.\ amplitude $\simeq 1\,\%\,$--\,$2\,\%$.  In some cases, the oscillations were observed to disappear immediately before the onset of a burst, which suggests a link to the thermonuclear burning on the neutron star envelopes.  These QPOs have been identified as a special mode of nuclear burning predicted by \citet{Paczynski1983} with a single zone model, where the oscillations are a consequence of the accretion-rate boundary between stable and unstable nuclear burning, and studied in detail in multi-zone models by \citet{heger2007}.  

With the stellar hydrodynamics code \textsc{Kepler} \citep{Weaver1978,Woosley2004,heger2007}, simulations are able to reproduce the oscillatory burning at accretion rates closed to the Eddington-limit, defined as $\dot{M}_{\text{Edd}}=1.75\times10^{-8}\times1.7/(X+1)\,\text{M}_{\odot}\,\text{yr}^{-1}$, where $X$ is the hydrogen fraction.  This finding was confirmed later by Modules for Experiments in Stellar Astrophysics (\textsc{MESA}; \citealt{mesa1,mesa2,mesa3}), e.g., first for marginally stable burning of helium accretion \citep{zamfir2014}.  However, one unsolved puzzle is that the accretion rate where QPOs are predicted by models is typically an order of magnitude higher than that from observations.

Observationally, such oscillations are rather rare.  So far, only eight sources, plus one disputed, have been found that show this feature \citep{tse2021,4u1730_qpo}.  One of the reasons is because these oscillations are expected to be weak, with typically less than 2.5\% oscillation amplitude \citep{tse2021}.  Furthermore, the oscillations vary in both frequency and amplitude over time, making them difficult to detect.  These properties motivate the adoption of more advanced search techniques in an attempt to find additional examples of the signals in order to better understand the physics of the oscillatory burning.


\src \; ($l=340.8^\circ$, $b=-2.48^\circ$) is an exceptional transient LMXB having both ``Z'' and ``atoll'' features.  The classification refers to the pattern that sources trace out in the X-ray color-color diagrams (CD), and the majority of souces only show features of one class consistently \citep{z_atoll}.  \src\ was first detected on 2006 January 18 by the All Sky Monitor onboard \textsl{RXTE} at the commencement of a new accretion outburst \citep{2006ATel..696....1R}.  Extensive follow-up observations were made with PCA throughout the outburst, which spanned about 550 days.  \cite{h07}, \cite{l09} and \cite{h10} undertook detailed studies of the CD of the source over the observations.  They found that it traced a Z pattern almost the entire outburst, but later, when the persistent luminosity dropped near the end of the outburst, it showed atoll source features.  Three bursts with photospheric radius expansion (PRE) were detected during the decaying phase of the outburst, leading to an estimation of a distance to be $6.4\pm 0.1\,\mathrm{kpc}$ \citep{minbarpaper}.  \citet{j1701_khzqpos} also detected $\mathrm{kHz}$ QPOs from the source, similar to those observed in several other sources.  These are thought to be associated with the dynamics of matter orbiting the surface of a neutron star rather than nuclear burning \citep{vander2001}.  
 
The source commenced a new outburst on 2022 September 6, detected by the Gas Slit Camera onboard the Monitor of All-sky X-ray Image (\textsl{MAXI}/GSC \citealt{maxigsc}) instrument  \citep{j1701_new_outburst}.  This latest outburst  probably continued until March 2023 \citep{j1701_quiescent}.

In this paper, we report the discovery of $\mathrm{mHz}$ QPOs in \src \; from a search of the entire archival \textsl{RXTE}/PCA data set using the Multi-INstrument Burst ARchive (MINBAR \citealt{minbarpaper}).  In \S\ref{methods} we demonstrate the methodology and search procedure in detail.  In \S\ref{results} we show our results and also explain the reasons for supporting the detected $\mathrm{mHz}$ QPOs are caused by the marginally stable burning.
Finally, in \S\ref{discussion} we examine the new discovery in the context of the other sources, as well as the quasi-stable nuclear burning paradigm.

\section{Observations, data analysis and noise simulation}
\label{methods}

As part of a more comprehensive search, we analysed all archival observations of \src \; collected with the Proportional Counter Array (PCA) on board \textsl{RXTE} that are available within the MINBAR sample \citep{minbarpaper}. 
 First, to avoid any background contamination, we only chose observations with flag \texttt{-}, \texttt{a}, \texttt{d}, or \texttt{f}, as described in Table 12 from MINBAR, to rule out observations with significant flux contributed by other sources in the field of view.  As we focus on signals in the frequency range from $5$--$15\;\mathrm{mHz}$, we searched the Standard-2 mode $16$-s binned source light curves from $2$ to $9\,\mathrm{keV}$, which have sufficient time resolution and cover the dominant energy range for $\mathrm{mHz}$ QPOs \citep{2008ApJ...673L..35A}.  We chose for our analysis the standard product source lightcurves, which were not background-subtracted. This was because the background-subtracted lightcurves in some observations from the observation standard products, available through the High-Energy Astrophysics Science Archive Research Center (HEASARC), were missing.  We found that the differences between the results from both are negligible.  We ignored around 4.2\% of the observations that have negative count rates in their light curves.

\subsection{Wavelet analysis}
Here we describe the application of the continuous wavelet transform (CWT) to the analysis of time-series with varying frequency signals.  This technique has been applied to discover non-stationary oscillations, and its detailed explanation can be found in \citet[hereafter TC98]{tc98}.  Another method, the windowed Fourier transform, analyses the signal by sliding a window of length $T$ across a time series at each time step.  For the time series with timing resolution $\delta t$, each windowed transform is computed for all frequencies from $T^{-1}$ to $(2\delta t)^{-1}$, which is inefficient as it requires extra computational time for frequencies outside the range of interest.  Wavelet analysis, however, is more flexible in terms of the choice of frequency range.  

\begin{figure}
    \centering
	\includegraphics[width=\columnwidth]{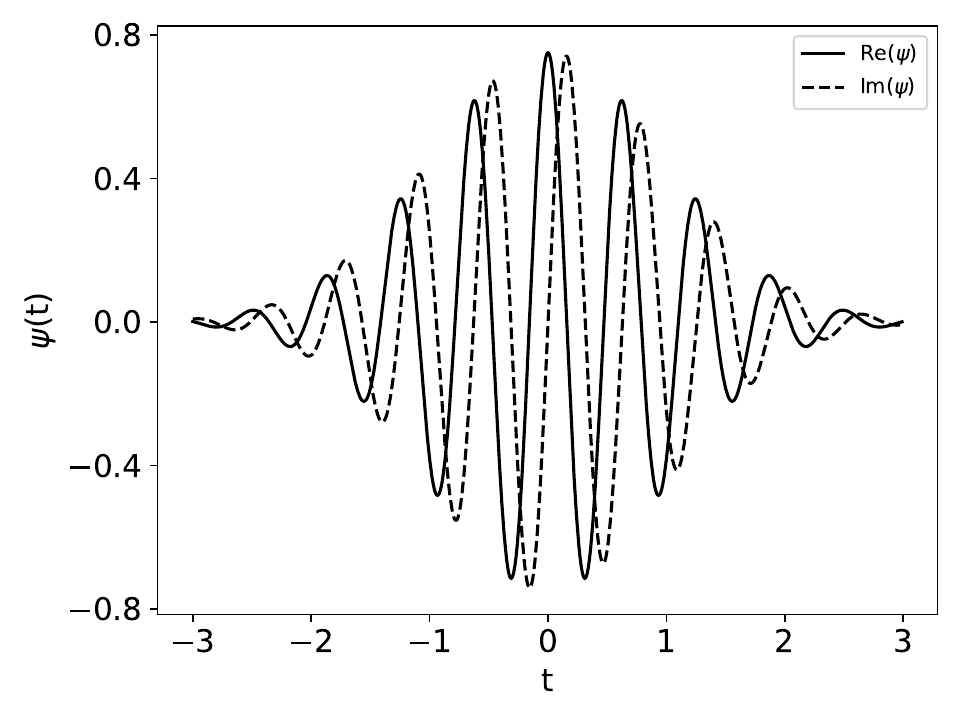}
    \caption{The real (\textsl{solid line}) and imaginary (\textsl{dashed line}) parts of a Morlet wavelet with $\omega_0=10$, which was adapted as the wavelet function in our analysis.}
    \label{fig:morlet}
\end{figure}

A fundamental concept in CWT is the wavelet functions.  The wavelet analysis makes use of functions which have zero mean and are localised in time and frequency space.  Depending on the subject to be studied, there are several standard functions that are used for CWT.  One commonly used function is a Morlet wavelet, which we also adopted in this work.  The Morlet wavelet consists of a sinusoidal waveform modulated by a Gaussian envelope over time $t$:
\begin{align}
    \psi_{\text{Morlet}}(t) = \pi^{-1/4}e^{i\omega_0t}e^{-t^2/2}\;,
\end{align}
where $\omega_0$ is the dimensionless frequency that controls the number of waves within the envelope.  For a Morlet wavelet with the same $\omega_0$ but with a more extended envelope, its transform returns information with higher frequency resolution but lower time resolution.  This trade-off is also a benefit in using wavelet analysis to search for quasi-periodic signals.  We chose $\omega_0=10$ in our study to ensure that both resolutions are balanced (see Figure~\ref{fig:morlet} for the example of the Morlet wavelet with $\omega_0=10$).  The CWT of a discrete time-series, $x_k$, of length $N$ can be denoted as
\begin{align}
W_k(s) = \sqrt{\frac{\Delta t}{s}}\;\sum_{k^{\prime}=0}^{N-1}x_k^{\prime} \psi^*\left[\frac{(k^{\prime} - k)\Delta t}{s} \right]\;,
\end{align}
where $\Delta t$ is the resolution of the time-series; $s$ is the wavelet scale, which sets the corresponding frequency for the transform; $k$ is the localised time index; and $\sqrt{\Delta t/s}$ is the normalisation factor such that the wavelet transform is weighted only by the amplitude of the Fourier coefficients $x_k^{\prime}$ instead of the wavelet function which depends on the scale.  Therefore, a dynamical power spectrum can be constructed showing the evolution of amplitude, defined as $|W_k(s)|^2$, over time and scale (frequency) domains.  We used the Python module PyCWT\footnote{PyCWT: spectral analysis using wavelets in Python; \url{https://pypi.org/project/pycwt/0.3.0a22/}} to perform the wavelet transforms, which are done in Fourier space for efficiency.  By the convolution theorem, the wavelet transform can also be performed by the inverse Fourier transform of the product of Fourier transforms of $x_k$ and $W_k(s)$, i.e., 
\begin{equation}
W_k(s) = \sqrt{\frac{2\pi s}{\Delta t}}\;\sum_{j=0}^{N-1}\hat{x}_j\hat{\psi}^*(s\omega_j)e^{i\omega_jk\Delta t}\;.
\end{equation}

\subsection{Data processing and analysis}
\label{subsec:analysis}
For every {\it RXTE}/PCA observation from MINBAR, we first excluded the region surrounding any bursts, from $5\,\mathrm{s}$ prior to the burst start, through four times of the burst duration afterwards.  This ensures that the flux from the burst has dropped back to the persistent level in each observation.  Then, we subdivided the light curves into individual segments by gaps from the burst clearance, or caused by any orbital shutdown of the telescope.  Each segment was treated independently for the subsequent analysis.  

We first measured the broad-band power spectrum of each observation segment by taking the Fourier transform with Leahy normalisation \citep{leahy1983AL},
\begin{align}
    P_i = \frac{2|a_i|^2}{N_\gamma}\;,\quad\;a_i = \sum_{k=0}^{N-1}x_ke^{i\omega_it_k}
    \;,
\end{align}
where $N_\gamma$ is the total number of photons; $x_k$ is the time-series; the complex number $a_i$ is the Fourier amplitude, to obtain the power spectrum averaged over the entire observation.  The spectrum may be contaminated by ``red noise'', which contributes to broad-band power in the low frequency range \citep[e.g.,][]{van_der_1988}, and hence it is more difficult to detect and estimate the significance of mHz QPOs features.  To address its potential impacts on both Fourier and wavelet spectra, we modelled the underlying red noise following the approach from \cite{Vaughan05}.  We excluded the frequency range of interest from 5--15$\,\mathrm{mHz}$ (for most of the signals falling within this frequency range \citealt{tse2021}) and modelled the Fourier spectrum with a power-law model 
\begin{align}
L_i = Af^{-\alpha}_i + C\;,
\label{noise_model}
\end{align}
where $f_i$ and $L_i$ are frequency and the modelled noise power of the spectrum, respectively, and $A$, $\alpha$, and $C$ are the normalisation, the power-law index, and additive constant of the model parameters, respectively.  After fitting the red noise model, we normalised the power spectrum by the model and divide it by $2$, i.e., $2P_i/L_i$, such that the powers become distributed as $\chi^{2}_2$ over the spectrum and are independent of frequency.  

Prior to taking the CWT for the individual segments of light curves, we detrended the light curve by subtracting a best-fit quadratic curve.  Finally, we took the CWT for each segment and normalised the wavelet spectrum by their corresponding best-fit red noise model.

To better assess the distribution of powers for each wavelet spectrum, we applied a Monte Carlo approach to generate realisations of each distribution of wavelet powers, and defined the detection threshold based on the simulations.  The idea was inspired by \cite{Bilous_2019}, where they applied the approach to search for burst oscillations in \textsl{RXTE} data.  For each segment of the light curve, we first generated synthetic data sets using the software package \textsc{stingray} \citep{stingray} based on the mean, fractional R.M.S, as well as the red noise model of the Fourier spectrum of the original time-series.  We then followed the same procedure as performed on the real data to create a wavelet dynamic power spectrum for each synthetic data set, i.e., fitting the red noise model, detrending the segment, and performing CWT including the red noise renormalisation.  By combining the simulated wavelet powers from all timing and frequency bins in the synthetic spectra, we defined the detection threshold for each individual segment as the $n$th largest power, where $n$ is the number of simulated time-series.  Our goal by choosing this threshold is to achieve a false-positive rate of 1 per each observation segment, on average.  Since the maximum power reached in each synthetic power spectrum can be quite variable, we found that, generally, the threshold converged approximately when $n \ge 20$, with more than 3 million synthetic wavelet powers in each combined distribution.  We took $n=60$ in all cases for the search to obtain more stringent detection thresholds.
  
\begin{figure*}
    \centering
	\includegraphics[width=\textwidth]{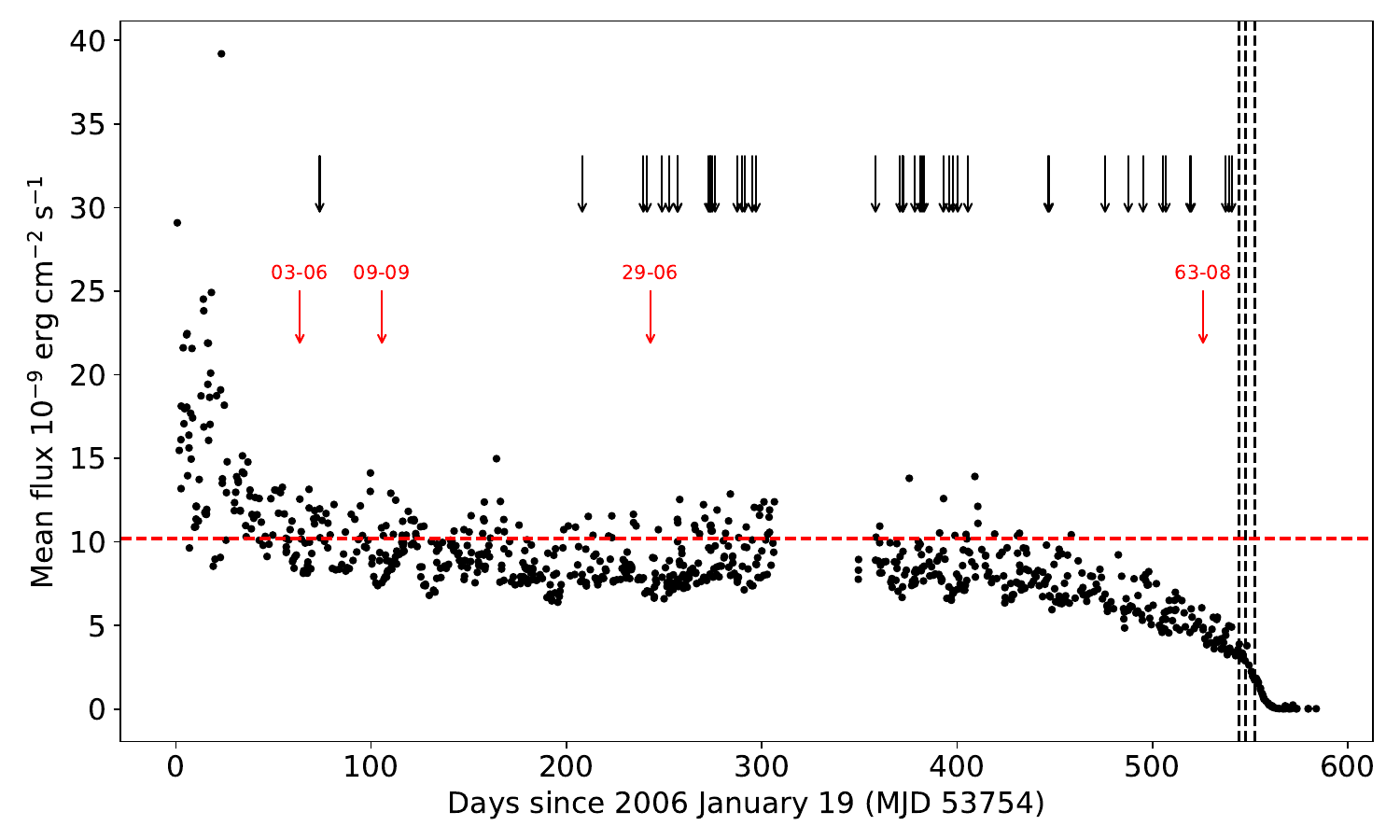}
    \caption{\textsl{RXTE}/PCA observations of \src \;, showing the persistent 3--25~keV flux.  The red arrows indicate the four most significant candidates with the detected QPOs, labeled by the last four digits of their observation IDs.  The remaining 43 lower-significance detections are indicated by black arrows.  The times of the three bursts observed by \textsl{RXTE} are indicated by dashed vertical lines (see Section~\ref{results}), whereas the dashed horizontal line in red color represents the 3--25~keV luminosity level at $5 \times 10^{37}\;\mathrm{erg}\;\mathrm{s}^{-1}$ with a distance of $6.4\;\mathrm{kpc}.$}
    \label{fig:lc}
\end{figure*}

\section{Results}
\label{results}
\begin{figure*}
    \centering
    \begin{subfigure}[b]{0.475\textwidth}
         \centering
         \includegraphics[width=\textwidth]{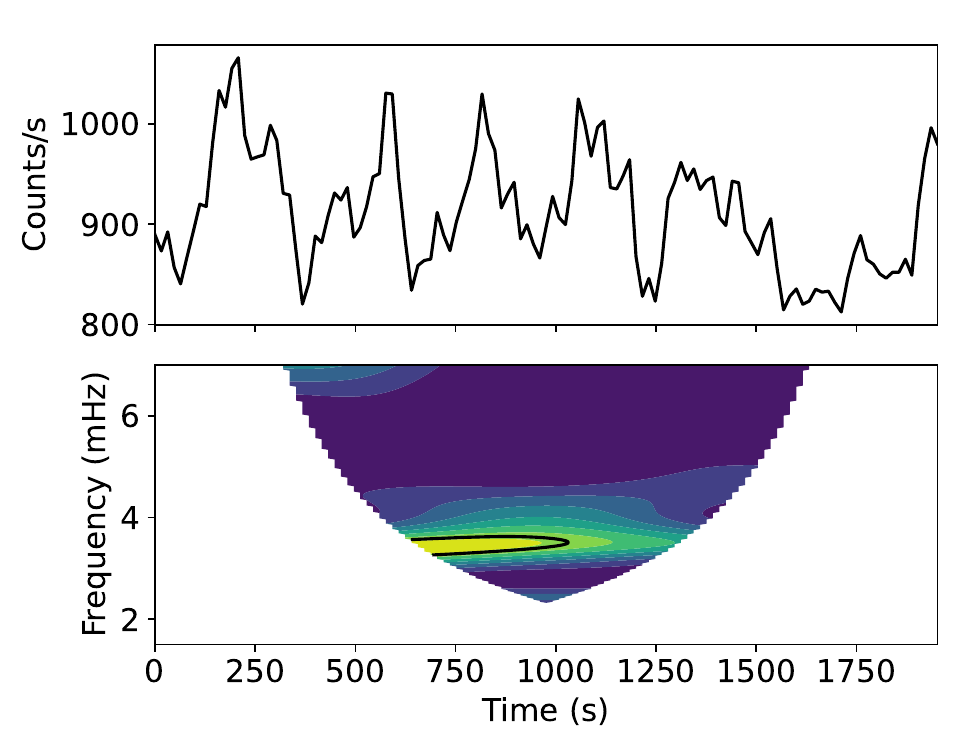}
         \caption{92405-01-03-06}
         \label{fig:ws_03-06}
     \end{subfigure}
    \begin{subfigure}[b]{0.475\textwidth}
         \centering
         \includegraphics[width=\textwidth]{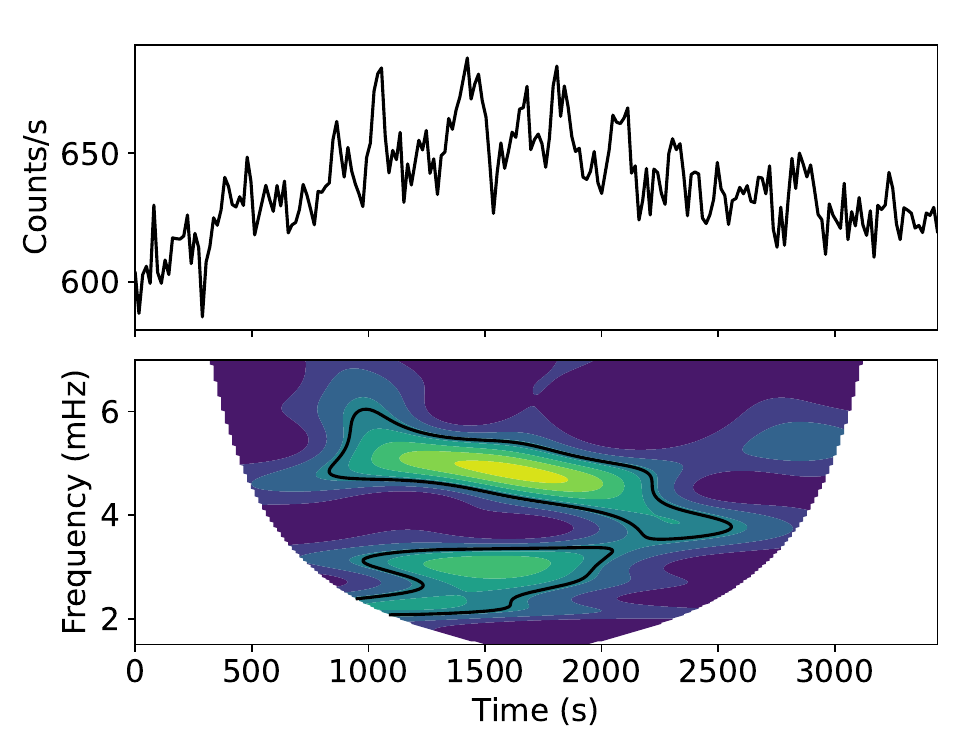}
         \caption{92405-01-09-09}
         \label{fig:ws_09-09}
     \end{subfigure}
    \begin{subfigure}[b]{0.475\textwidth}
         \centering
         \includegraphics[width=\textwidth]{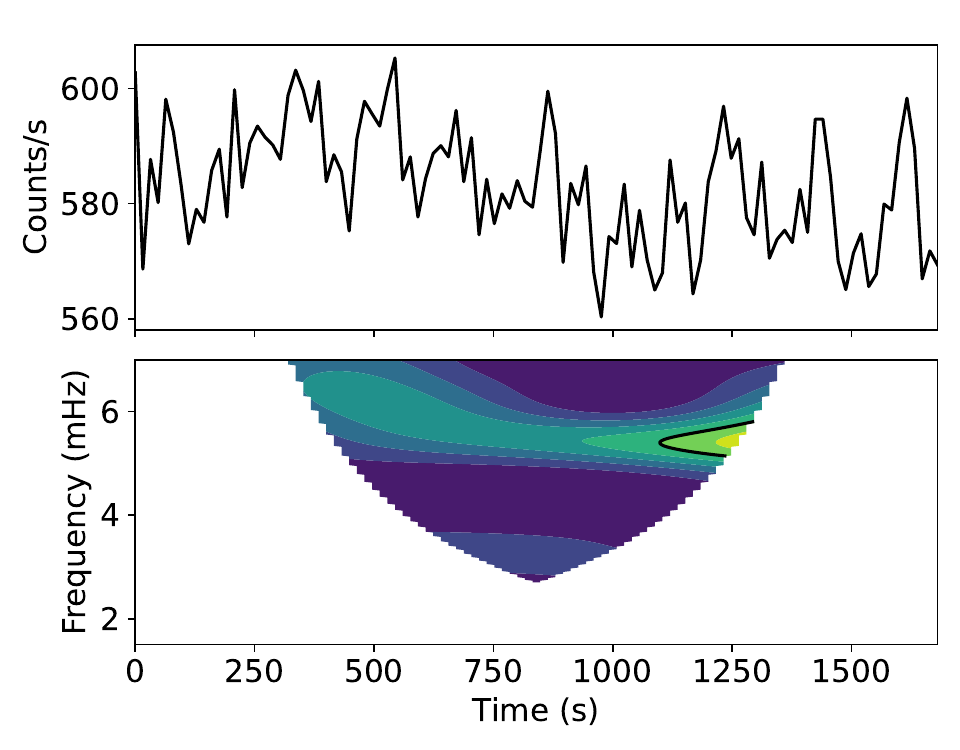}
         \caption{92405-01-29-06}
         \label{fig:ws_29-06}
     \end{subfigure}
    \begin{subfigure}[b]{0.475\textwidth}
         \centering
         \includegraphics[width=\textwidth]{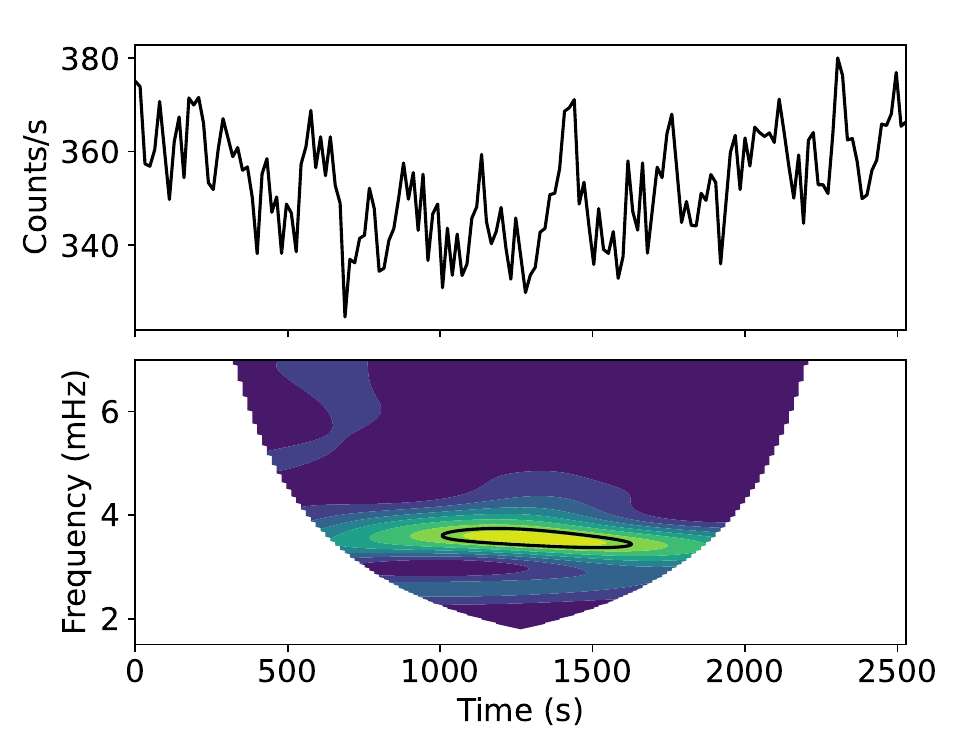}
         \caption{92405-01-63-08}
         \label{fig:ws_63-08}
     \end{subfigure}
     \caption{X-ray light curve ({\it top panels}) and wavelet spectrum ({\it bottom panels}) for the four observations with detected $\mathrm{mHz}$ QPOs.  The 16 seconds binned light curves cover an energy range from 2 to $9\;\mathrm{keV}$.  The colour scale was normalised by the detection threshold defined in Section~\ref{subsec:analysis} for each spectrum individually, and powers above the thresholds are represented by the black contours.  The white areas in the lower panels indicate where the wavelet powers drop by a factor $e^{-2}$ because of the edge effect from finite time-series \citep{tc98}.  The resulting spectra show signals that exceed the thresholds in the $\mathrm{mHz}$ domain.}
     \label{fig:ws}
\end{figure*}

We found that \src \; shows significant periodic signals at around $5\;\mathrm{mHz}$ in the wavelet spectra.  We studied the source again in more detail, but with background-subtracted lightcurves in this time.  We expanded the frequency range down to $1.5\;\mathrm{mHz}$, motivated by the excess power measured at the previous lower frequency limit, to see whether more signals are detectable with a broader frequency coverage.
The result shows 47, out of 860 observations (selected from Section~\ref{methods}) with significance above the detection thresholds.  Among these observations, we demonstrate the four strongest candidates (observation IDs 92405-01-03-06, 92405-01-09-09, 92405-01-29-06 and 92405-01-63-08) with each maximum power exceeding $4\sigma$ of the simulated wavelet power distribution.  The light curves and their corresponding wavelet spectrum are shown in Figure~\ref{fig:ws}.

\begin{figure*}
    \centering
    \begin{subfigure}{0.475\textwidth}
    \includegraphics[width=\columnwidth]{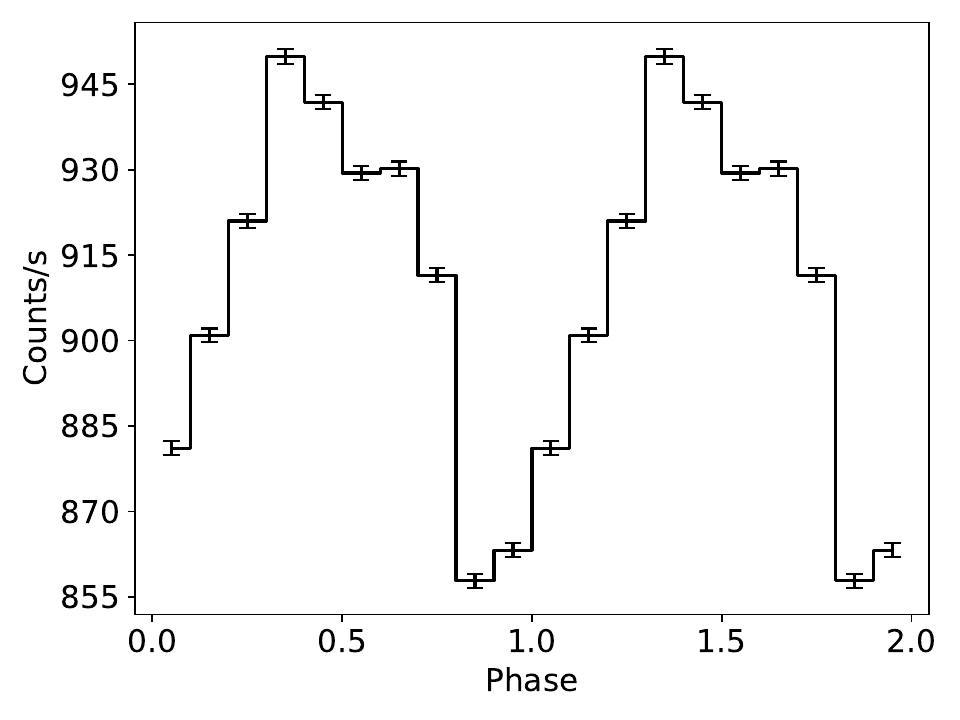}
    \caption{92405-01-03-06}
    \end{subfigure}
    \begin{subfigure}{0.475\textwidth}
    \includegraphics[width=\columnwidth]{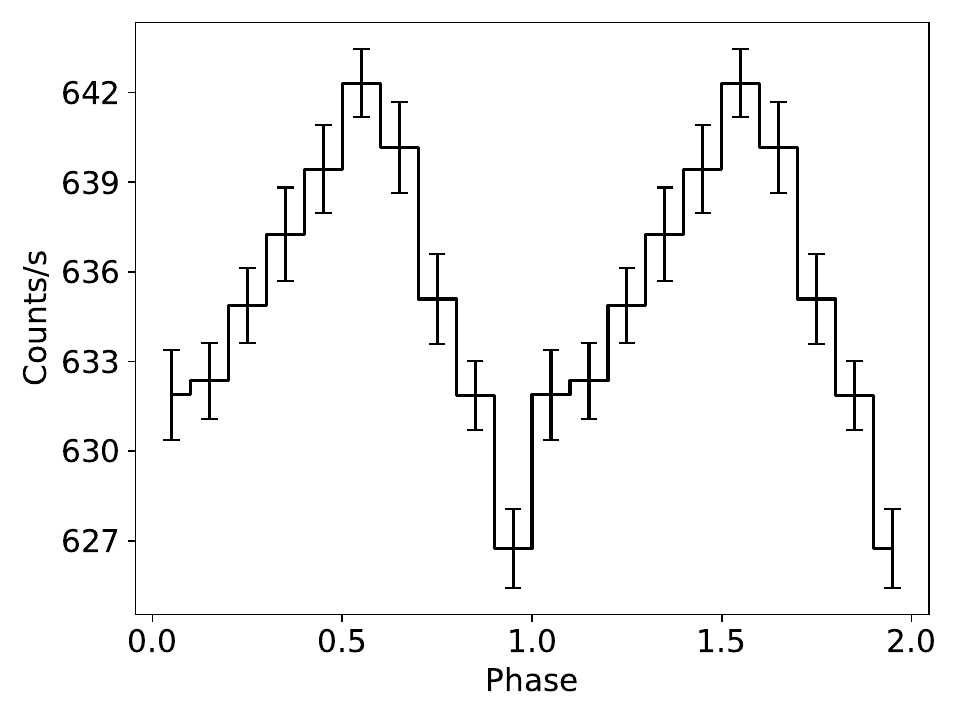}
    \caption{92405-01-09-09}
    \end{subfigure}
    \caption{The phase-folded $\mathrm{mHz}$ oscillation profiles from Observation 92405-01-03-06 and 92405-01-09-09.  The profiles are created from the 16-s binned light curves folding at frequencies $3.5$ and $4.8\;\mathrm{mHz}$, respectively.  Counts within the energy range 2--$9\;\mathrm{keV}$ were included.  The oscillation amplitudes are $3.54 \pm 0.04\,\%$ and $0.74 \pm 0.05\,\%$ respectively.}
    \label{fig:pulse}
\end{figure*}

The four observations were made on MJD 53817, 53859, 53996, and 54279 spanning almost the entirety of the outburst (see Figure~\ref{fig:lc}).  We summarise their QPO signal properties in Table~\ref{table}.  The CWT spectra show QPOs with powers that exceed the thresholds in $\sim 3.5$ to $5.6\;\mathrm{mHz}$.  We estimated the fractional RMS amplitudes of the detected $\mathrm{mHz}$ QPOs for the two observations by phase-folding each light curve at the peak frequency in the spectrum, resulting in amplitudes of $3.54\pm0.05\%, 0.74 \pm 0.05\%, 0.78 \pm 0.07\%$ and $1.39 \pm 0.08\%$, respectively.  Figure~\ref{fig:pulse} shows two oscillation profiles (for observation IDs 92405-01-03-06 and 92405-01-09-09).   

\begin{figure}
    \centering
    \includegraphics[width=\columnwidth]{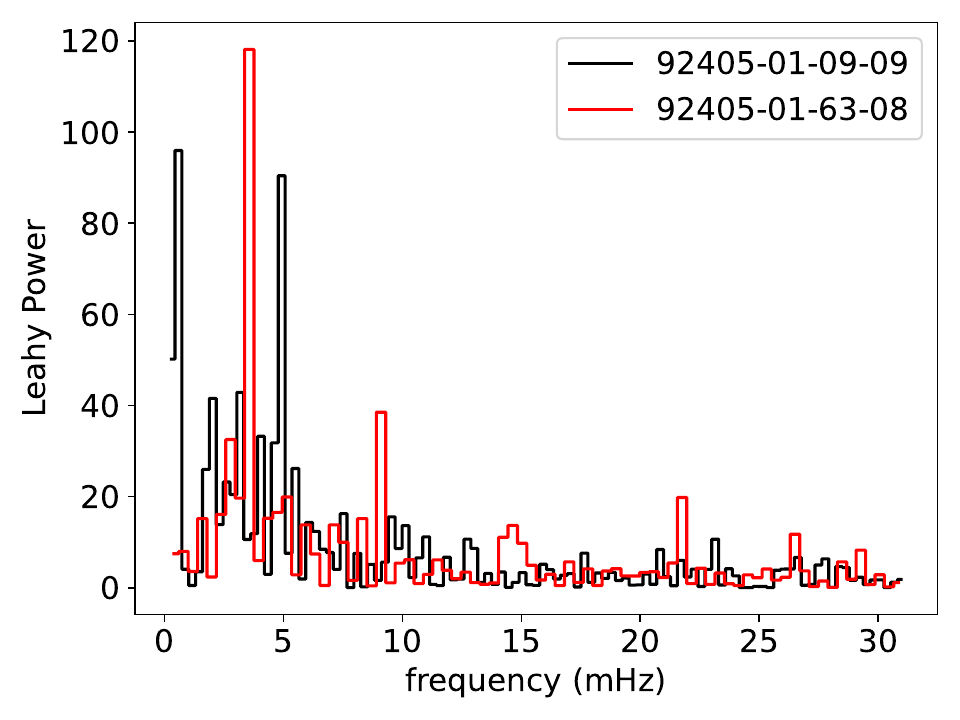}
    \caption{Combined Fourier spectrum from Observation 92405-01-09-09 and 92405-01-63-08 with Leahy normalisation. Strong peaks can be found at $\sim 5$ and $3.5\;\mathrm{mHz}$ in the spectra, respectively.}
    \label{fig:Fourier}
\end{figure}

Figure~\ref{fig:Fourier} shows the Fourier power spectra of the detrended light curves from observations 92405-01-09-09 and 92405-01-63-08, demonstrating that the detected signals are not merely an artefact of the wavelet algorithm.  Both power spectra show strong peaks at frequencies where their corresponding wavelet spectrum also shows significant signals.

Three bursts were detected by \textsl{RXTE}/PCA from \src\ on MJD 54298, 54301, and 54306 (MINBAR IDs \#3567, 3569 and 3572, respectively; Figure~\ref{fig:lc}).  We ignore a weakly significant burst candidate detected by \textsl{INTEGRAL}/JEM-X on MJD 54176 (MINBAR ID \#6328 \citealt{minbarpaper}) as it may arise from an accretion rate fluctuation.  \citep{h10} proposed that the first burst was in a transition phase from Z to atoll, whereas the other two were in the atoll phase.  We did not find any evidence for QPOs immediately before or after these bursts, unlike the disappearance of signals before the onset of bursts, such as seen in previous reported QPOs (e.g., \citealt{firstqpo,4u1730_qpo}).

\begin{table*}
	\centering
	\caption{Key figures and properties of $\mathrm{mHz}$ QPOs from from \src \;.}
    \label{table}
    \begin{threeparttable}
    \begin{tabular}{lccccc}
    \hline
    ObsID & Time (MJD) & Total exposure (s) & \makecell{Accretion luminosity\\($10^{37}\,\mathrm{erg}\,\mathrm{s}^{-1}$)} & \makecell{QPOs frequency\\$(\mathrm{mHz})$} & \makecell{Max R.M.S\\$(\%)$} \\
    \hline
    92405-01-03-06 & 53817 & 1968 & 8.86$\pm 0.08$ & 3.5 & $3.54 \pm 0.04$ \\
    92405-01-09-09 & 53859 & 3456 & 5.54$\pm 0.07$ & 4.8 & $0.74 \pm 0.05$\\
    92405-01-29-06 & 53996 & 1696 & 4.94$\pm 0.07$ & 5.6 & $0.78 \pm 0.07$ \\
    92405-01-63-08 & 54279 & 2528 & 3.36$\pm 0.07$ & 3.5 & $1.39 \pm 0.08$ \\
    \hline
    \end{tabular}
    \begin{tablenotes}[flushleft]
    \item We adopted the bolometric correction $1.44 \pm 0.07$ and the distance $6.4\;\mathrm{kpc}$ from \cite{minbarpaper} for the estimations of luminosity, and we did not take into account the uncertainties in the distance.
    \end{tablenotes}
    \end{threeparttable}
\end{table*}

\section{Discussion and Conclusion}
\label{discussion}
We report the discovery of $\mathrm{mHz}$ QPOs in the LMXB \src \; as a result of a broader search of \textsl{RXTE}/PCA observations from bursting sources.
We found that 47 out of 860 observations during the 2006--7 outburst exhibited signals above the detection thresholds, defined by a Monte Carlo approach, in their wavelet spectra.  The number of detections is less than our expectation, as our approach should pick up a false-positive rate of 1 per each observation segment, on average.  However, it may suggest that our approach to defining the detection threshold is rather conservative.  This idea is further reinforced by the fact that our approach detects the QPOs in 67 out of 1442 observations of 4U~1636$-$536 made by PCA, where the ratio is lower than the study from \cite{Lyu_2016} with 207 detections out of more than 1,500 observations.  Therefore, there is the possibility that the oscillations present in other \src\; observations, but being overlooked due to the strict detection thresholds we defined with our approach.

Among the $47$ observations with detected $\mathrm{mHz}$ QPOs, we show the properties of the four most significant candidates with each maximum power exceeding $4\sigma$ of the simulated wavelet power distribution.  The signals from these four observations occurred when the accretion luminosity was about $0.16$ -- $0.43\,L_{\text{Edd}}$ ($L_{\text{Edd}}=2.08 \times 10^{38}\;\mathrm{erg}\;\mathrm{s}^{-1}$ for accreting solar composition).  The relatively low oscillatory amplitudes of between 0.74--3.54\% RMS indicate that oscillations are not likely caused by the "heartbeat model" found in another LMXB \citep{j00291_qpos}, which instead have amplitudes of typically 30\%, and are linked to the movement of the inner accretion disk \citep{2011ApJ...742L..17A}.  The source was in the ``Z'' phase during the four observations (and also the other 43 according to \citealt{l09}), in contrast to most of the previous reports in which oscillations were found in atoll sources.

\citet{tse2021} reported examples of mHz QPOs detected in 5 of 8 other sources that vanished immediately before a Type I X-ray burst (see their Table~1 for summary).  This coincidence is strong evidence for the thermonuclear origin of the oscillations.  We do not find a similar phenomenon in this source. 
The $\mathrm{mHz}$ QPOs found in \src \; where the oscillations disappear with the occurrence of bursts at lower accretion rates is explicable by 1D models of nuclear burning.  In theory, the stability boundary of nuclear burning occurs at a narrow range of accretion rates, which sits above the bursting regime \citep{heger2007,galloway_2021}.  Therefore, one would expect that the QPOs take place at close to the Eddington accretion rate, whereas bursts resulting from unstable burning occur at lower accretion rates.  Our results show that bursting in \src \; commenced  when the accretion rate dropped during the decaying tail of the outburst.  This behaviour suggests that the source has entered the unstable burning regime, where the accreting fuel is piled up and ignited episodically (resulting in bursts), instead of burnt steadily and quasi-stably.  This phenomenon is a better match to the theory \citep{Paczynski1983, heger2007}, rather than having both concurrently at the same level of accretion rate. 

The coexistence of bursts and $\mathrm{mHz}$ QPOs in the other sources is harder to reconcile with theory.  \citet{Cavecchi2017} suggested that nuclear burning depends on effective gravity, which is different across the latitudes of a neutron star due to the typically rapid (few hundred Hz) rotation rates.  The variation in effective gravity with latitude may lead to different burning regimes across the neutron star surface, and the possibility of stable burning coexisting with stable.  With the complexity of the actual nuclear burning in a three-dimensional environment, it is hard to conclude whether the $\mathrm{mHz}$ QPOs found in the case of coexistence with bursts and \src \; have the same origin or not.

Based on our results, we propose that the $\mathrm{mHz}$ QPOs found in \src \; are due to marginally stable nuclear burning.  If so, it will be the tenth source in which this special mode of nuclear burning has been detected, after 4U~1636$-$536, Aql~X$-$1, 4U~1608$-$52 \citep{firstqpo}, 4U~1323$-$619 \citep{2011ATel.3258....1S}, EXO~0748$-$676 \citep{exo_qpos}, IGR~J17480$-$2446 (\citealt{2012ApJ...748...82L}, although see \citealt{4u1730_qpo}), GS~1826$-$238 \citep{2018ApJ...865...63S}, 1RXS~J180408.9$-$342058 \citep{tse2021}, and the currently discovered in 4U~1730$-$22 \citep{4u1730_qpo}.  Moreover, it is the first source that exhibits the oscillations when following the Z pattern in the CD.  It implies that the occurrence of QPOs is based on the actual thermal state of the burning layers of the neutron star, on which the source state alone cannot completely reflect because the persistent spectrum contains emissions from the inner disk and the compact object \citep{van_der_Klis2006}.  Similarly, the presence or absence of QPOs does not solely depend on the instantaneous accretion rate (indicated by the X-ray flux), since the accretion history also has an impact on the thermal state of the burning layers.  The oscillations, however, are expected to occur when a source switches from stable burning to bursting regimes, similar to the case of \src \; in this report, or vice versa.




 



\section*{Acknowledgements}

Parts of this research were conducted by the Australian Research Council Centre of Excellence for Gravitational Wave Discovery (OzGrav), through project number CE170100004.
This work was supported in part by the National Science Foundation under Grant No. PHY-1430152 (JINA Center for the Evolution of the Elements).
This research has made use of data obtained through the High Energy Astrophysics Science Archive Research Center Online Service, provided by the NASA/Goddard Space Flight Center.
We thank Celia Sanchez-Fernandez for the analysis of \textsl{INTEGRAL}/JEM-X data. 
\section*{Data Availability}
The \textsl{RXTE} data underlying this article are available in the High Energy Astrophysics Science Archive Research Center (HEASARC), at (\url{https://heasarc.gsfc.nasa.gov}).
 



\bibliographystyle{mnras}
\bibliography{ms} 





\bsp	
\label{lastpage}
\end{document}